# Occultations of HIP and UCAC2 stars downto 15$^m$ by large TNO in 2004-2014

© 2004 D.V. Denissenko

*Space Research Institute (IKI), Moscow, Russia*

Occultations of stars brighter than 15$^m$ by largest TNOs are predicted. Search was performed using the following catalogues: Hipparcos; Tycho2 with coordinates of 2838666 stars taken from UCAC2 (Herald, 2003); UCAC2 (Zacharias et al., 2003) with 16356096 stars between 12.00 and 14.99 mag to the north from -45 declination. Predictions were made for 17 largest numbered transneptunian asteroids, recently discovered 2004 DW and 4 known binary Kuiper Belt objects. 64 events occuring at solar elongation of 30° and more are selected, including exceptionally rare occultation of 6.5$^m$ star by double (66652) 1999 RZ$_{253}$ on 2007 October 4th. Observations of these events by all available means are extremely important since they can provide unique information about the size of TNOs and improve their orbits dramatically.

## INTRODUCTION

Over 800 Transneptunian Objects (TNO) are discovered since 1992 with 67 of them being numbered and 7 having proper names as of February 2004. No reliable measurements of their sizes have been obtained so far. Angular sizes even of the largest objects are about 0.04"±0.01" (Quaoar) which is at the limit of Hubble Space Telescope resolution. Most size estimates are based on the indirect methods strongly dependant on assumptions. Albedoes 0.04-0.08 are generally supposed for TNOs, but they are definitely varying among different objects.

The only direct method to measure sizes of single TNOs at present is observing stellar occultations by them. Over 500 occultations of stars by Main Belt asteroids have been observed so far, with 17% of them having 5 or more chords measured. For some objects size and shape were determined with an equivalent angular resolution of 0.002".

## OCCULTATIONS BY TNO

TNOs are typically at 30-50 AU distances from the Earth corresponding to 0.2"-0.3" parallax. This means that occultation will only happen somewhere on the Earth if the geocentric path of the object passes within 200-300 mas from the star. Combined with a very slow angular motion of TNO (0.1-1 mas per second of time) it makes rate of stellar occultations by them about three orders of magnitude smaller than by main belt asteroids and about 100 times less than that of Jovian Trojans, approximately inversely proportional to $(a-1)^2$.

Estimates show that any given TNO will occult a star brighter than 15$^m$ on average approximately every 4 or 5 years. For ~10$^5$ HIPPARCOS stars brighter than 10-11$^m$ we should be expecting one occultation in 10 years by 20 largest transneptunian objects, and only one event per century for ~10$^4$ stars brighter than 6.5$^m$.

## STAR CATALOGUES ACCURACY

Error in coordinates of occulted star will significantly increase the uncertainty of both event time and position. At 40 AU distance, 1 milliarcsecond is corresponding to 30 km. 0.25" error in star coordinates will shift the occultation path by ~1 Earth radius. Since the positions of Transneptunian objects are known to 0.25" uncertainty, 0.05" accuracy of the star coordinates will be acceptable. Before 2003, coordinates were known to milliarcecond accuracy for ~100000 stars in HIPPARCOS catalogue. Tycho2 catalogue including ~2 Mln stars usually brighter than $12^m$ is essentially a by-product of the same HIPPARCOS project. It has systematic and statistical errors typically within 0.1"-0.2" but reaching 0.5"-1.0" in some fields. Thus Tycho2 is not satisfactory for predicting occultations by TNO.

Situation in astrometry was revolutionized in 2003 with the release of $2^{nd}$ USNO CCD Astrograph Catalogue UCAC2 (Zacharias et al.) with coordinates and proper motions for 48 Mln stars between $8^m$ and $16^m$. Position errors of $12^m$ stars in UCAC2 are 15-20 mas and 40-50 mas for $15^m$. This catalogue does not cover areas to the north from +45..+52 declination, but that is not a problem since no objects involved in current search will reach ±45 declination during next 10 years.

## PREDICTIONS FOR 2004-2014

Occultations from 2004 Jan 27 to 2014 Dec 31 were computed for the following objects:

a) Recently discovered largest known TNO 2004 DW with absolute mag $H_0$=2.2-2.4;

b) 17 numbered objects with $H_0$<5.5

2.6 (50000) Quaoar
3.2 (28978) Ixion
3.3 (55565) 2002 $AW_{197}$
3.3 (55636) 2002 $TX_{300}$
3.6 (55637) 2002 $UX_{25}$
3.7 (20000) Varuna
4.2 (42301) 2001 $UR_{163}$
4.5 (19308) 1996 $TO_{66}$
4.7 (26375) 1999 $DE_9$
4.7 (38628) Huya
4.8 (24835) 1995 $SM_{55}$
4.9 (19521) Chaos
4.9 (47171) 1999 $TC_{36}$ (binary)
5.2 (26181) 1996 $GQ_{21}$
5.3 (55638) 2002 $VE_{95}$
5.4 (15874) 1996 $TL_{66}$
5.4 (48639) 1995 $TL_8$

c) 3 numbered binary TNO

5.8 (26308) 1998 $SM_{165}$
5.9 (66652) 1999 $RZ_{253}$
6.6 (58534) 1997 $CQ_{29}$

Single body diameters were calculated from $H_0$ by the formula log D[km]=3.52-0.2*$H_0$. This corresponds to assumed albedo of 0.16 and is a conservative estimate. For (50000) Quaoar this formula gives 1000 km which is in a good agreement with the lower limit from Hubble direct imaging (1260±300 km). If albedo is 0.04, actual size of TNO will be exactly twice the value listed in table. For binary asteroids with two bodies of the same size and albedo diameter of each component will be 0.71*D.

Orbits of selected TNO from 2004 Jan. 27th version of astorb.dat (Bowell, 2004) were integrated through the end of 2014 and all geocentric conjunctions within 0.5" to HIP and UCAC2 stars were found. For the year 2004 occultations by 36 largest unnumbered objects with $H_0$<5.5 were also computed and 9 events have been found. Mean value of current ephemeris uncertainty (CEU) was 0.26" for the numbered objects and 1.0" for unnumbered ones.

Results of the search are presented in table. Intervals of time (rounded to 5 minutes) are given when occultation may happen on the night part of Earth with 0.25" uncertainty at both sides already taken into account. Due to the different speed of TNOs relative to Earth (from 25 km/s at opposition to 5 km/s and even less near the stationary point) these interval may vary from ~10 minutes to 1.5 hour in case of Varuna occultation in 2012. For double asteroids extra 0.25" were added since orbital distances in these pairs are also estimated as 6-8 thousand km.

Exceptional occultation of $6.5^m$ K0 star HIP 111398 = HD 213780 = SAO 146161 = ZC 3313 in Aquarius by binary object (66652) 1999 $RZ_{253}$ in 2007 should be noted. Additional observations of this TNO on the largest telescopes are extremely important to predict the separation and positional angle of components at the epoch of the event. International campaign involving observers in Australia, New Zealand, China, Japan, eastern Russia and Hawaii should be organized to measure the sizes and shape of this binary system.

Finder charts and preliminary path plots for occultations listed here are available at http://hea.iki.rssi.ru/~denis/TNOocc.html or by E-mail request to denis@hea.iki.rssi.ru.

OBSERVING METHODS

Observations can be performed by several techniques and different kinds of equipment. In principle, highest temporal resolution is not necessary for occultations by TNO. With event durations about several dozen seconds and even 2-3 minutes in some cases, 1-3 sec integration time will already give the unprecedented chord measurement accuracy. This is why one can use CCD imaging, CCD in drift-scan mode, photometers, high-sensitivity videocameras and even visual observations with stopwatch. CCD readout time should be reduced to minimum by binning pixels and exposing small window instead of full frame. Optimal exposure time should be selected depending on the magnitude of occulted star and detector sensitivity. Precise absolute timing of star disappearance and reappearance is not critical since with 5-10 minutes uncertainty of event time even 1-second accuracy will improve the orbital position by two orders of magnitude. For bright occultations fast photometers can provide the measurements of occulted star diameter at unprecedented angular resolution. Since TNOs are moving at 0.1-1 mas/sec rate (500-5000 times slower than the Moon), new close binaries can be discovered which are impossible to resolve by any other methods.

DISCUSSION

Search was intentionally limited to the largest objects with ephemeris uncertainty of 0.3" or better. Although smaller objects might (and definitely will!) occult brighter stars, detection probability of those events would be much smaller compared to those listed here. It is preferable to concentrate on 2-5 "special" events each year. Success can only be achieved by the joint efforts planned well in advance. Worldwide observations by as many telescopes as possible are necessary. Prediction of events till 2014 will allow to make changes in tight observing schedules at the large scopes. Additional observations of selected TNOs to improve ephemeris and prediction accuracy are also encouraged. New large objects will definitely be found in the nearest future. Predictions of occultations by them will be posted at author's site together with updates on those listed here.

## APPLICATIONS

Positive observation of stellar occultation by TNO will make significant impact in several areas of astronomy at once. It will allow to:
- measure the size and shape of these distant objects directly which is impossible by other means;
- give indications on their composition, and together with estimates of masses in binary systems directly determine their density;
- improve orbital position by 10-50 times assisting possible space missions to TNO;
- provide unique information valuable for understanding the theory of Solar system evolution;
- discover probable close binary objects with the separations between components of the order of 1000 km.

## ACKNOWLEDGEMENTS

I would like to thank Alexander Yascovich for lending the copies of UCAC2 catalogue and Andrei Plekhanov for writing LinOccult program which was used to plot the maps of occultation paths posted on the website.

OCCULTATIONS OF STARS BY TNO 2004-2014

```
Date of event   Time          Asteroid     size    Dur    Star              R.A. (2000.0) Dec
Year Mon DD     (UT)          name         km      sec    Name              hh mm ss.sss  dd mm ss.ss   Mag
2004 Jan 29     23:35-23:55   2003 AZ84    568     23.4   2UCAC 36806754    07 12 19.883  +14 04 20.02  14.5
2004 Feb 14     22:10-22:35   2001 XR254   255     14.2   TYC 1343-00785-1  06 54 14.687  +21 37 48.10  11.6
2004 Mar 07     03:15-03:30   2000 CO105   268     14.7   2UCAC 39141168    08 23 16.013  +20 33 01.42  12.3
2004 Mar 08     02:10-02:25   1999 CC158   241     11.6   2UCAC 40185292    09 15 58.264  +23 38 56.76  13.7
2004 Mar 20     13:35-13:45   1996 TL66    275     10.5   2UCAC 36232084    03 00 28.306  +12 40 40.89  13.4
2004 Mar 27     00:20-00:40   2002 KW14    302     21.2   2UCAC 24855750    15 34 29.721  -18 18 28.41  13.6
2004 Apr 07     10:30-11:00   2001 KD77    252     24.7   2UCAC 23658681    16 58 51.823  -20 36 51.67  11.7
2004 Jun 26     19:55-20:05   Chaos        347     11.6   2UCAC 40137180    04 08 53.436  +23 56 34.41  13.2
2004 Jun 29     18:25-19:10   1997 CQ29    158      9.5   2UCAC 34467578    11 04 07.504  +07 39 54.56  14.5
2004 Aug 08     23:15-23:30   2001 QD298   257     10.5   2UCAC 24989527    21 52 30.098  -18 08 13.32  14.5
2004 Nov 15     09:00-09:25   2000 CN105   340     32.5   2UCAC 35937214    10 34 08.639  +11 59 33.61  14.3
2005 Mar 13     21:25-21:35   2002 UX25    631     21.6   2UCAC 34953654    01 37 42.194  +09 22 06.65  13.2
2005 May 07     01:05-01:40   1999 RZ253   219     14.9   2UCAC 28298467    22 29 21.656  -10 02 41.04  14.9
2005 Sep 29     07:50-08:10   1995 SM55    363     16.6   2UCAC 39255131    02 07 19.531  +21 06 25.09  14.1
2005 Oct 24     08:05-08:30   1998 SM165   229      9.4   2UCAC 33162437    01 31 09.061  +04 23 37.99  14.4
2005 Dec 27     09:15-09:35   Huya         380     17.9   2UCAC 30864992    14 20 27.168  -02 54 42.58  11.7
2006 Jun 18     20:25-20:50   1999 KR16    229     17.4   2UCAC 28922119    14 18 38.914  -08 29 05.17  13.7
2006 Jul 26     07:55-08:25   Ixion        759     55.9   2UCAC 22945926    16 34 14.971  -22 04 37.89  13.7
2006 Oct 07     19:30-19:45   Ixion        759     36.0   2UCAC 22946030    16 35 05.045  -22 11 08.86  13.8
2007 Aug 04     04:20-04:35   2002 VE95    288     15.8   2UCAC 35331886    04 14 47.942  +10 08 18.19  13.0
2007 Oct 04     11:10-11:50   1999 RZ253   219     11.2   HIP 111398        22 34 06.604  -09 36 30.64   6.5
2008 Jan 22     06:55-07:30   Chaos        347     26.0   2UCAC 40810575    04 21 06.919  +25 33 26.70  10.8
2008 Feb 11     04:20-04:40   Varuna       603     28.8   TYC 1913-00670-1  07 18 50.100  +25 43 19.31  11.3
2008 Sep 10     12:25-13:15   Quaoar       1000   149.5   2UCAC 26249129    17 05 49.913  -15 20 44.26  13.7
2008 Sep 30     07:20-07:40   Ixion        759     44.6   2UCAC 22496420    16 44 05.610  -23 18 07.11  13.6
2008 Oct 07     19:40-19:55   Ixion        759     37.3   2UCAC 22496543    16 44 33.933  -23 19 19.45  12.8
2008 Dec 07     01:55-02:20   Varuna       603     29.0   2UCAC 40846256    07 29 48.171  +25 40 07.71  14.8
2009 Mar 10     08:25-08:40   2002 VE95    288     19.7   2UCAC 34960757    04 17 00.297  +09 18 45.63  13.3
2009 May 01     13:50-14:15   Quaoar       1000    54.1   2UCAC 26252549    17 18 12.542  -15 24 49.90  14.3
2009 Jul 18     04:40-04:50   2002 VE95    288     10.8   HIP   21308       04 34 30.274  +10 00 58.46   7.7
2009 Aug 22     10:40-11:15   2002 UX25    631     68.5   2UCAC 35326247    02 09 00.114  +10 07 56.40  14.1
2009 Sep 25     11:05-11:30   1995 TL8     275     15.1   TYC 1223-00579-1  02 40 23.728  +15 46 01.74  11.6
```

DENISSENKO D.V.

```
Date of event   Time           Asteroid    size   Dur    Star                R.A. (2000.0) Dec
Year Mon DD     (UT)           name        km     sec    Name              hh mm ss.sss  dd mm ss.ss   Mag
2009 Oct 09     10:15-10:35    2002 TX300  724    28.6   2UCAC 41650964    00 37 13.610  +28 22 22.98  13.1
2009 Oct 15     17:55-18:10    Quaoar      1000   45.8   2UCAC 26023237    17 12 33.132  -15 30 54.97  15.0
2010 Feb 19     22:55-23:20    Varuna      603    32.3   TYC 1914-00301-1  07 29 22.470  +26 07 23.23  11.0
2010 Jul 19     01:35-02:05    Ixion       759    42.4   2UCAC 22032118    16 54 25.862  -24 24 22.86  11.5
2011 Feb 22     02:25-02:50    1995 TL8    275    18.3   2UCAC 37301105    02 39 37.120  +15 41 48.75  13.6
2011 Jun 01     11:05-11:25    Ixion       759    30.3   2UCAC 21584537    17 03 39.446  -25 01 27.82  14.7
2011 Aug 03     22:35-23:10    1999 RZ253  219    10.8   2UCAC 29384710    22 59 38.963  -07 02 53.38  14.6
2011 Nov 04     00:35-00:50    Quaoar      1000   35.9   2UCAC 26027484    17 24 23.771  -15 38 43.34  14.6
2011 Dec 18     11:40-12:50    1999 TC36   347    40.7   2UCAC 32137136    01 35 51.911  +01 13 15.69  14.1
2012 Apr 16     04:20-04:55    Ixion       759    55.9   TYC 6816-00401-1  17 12 26.791  -25 34 10.85  10.8
2012 May 01     03:35-03:55    1999 RZ253  219    10.2   2UCAC 29759182    23 04 54.428  -06 29 06.50  14.9
2012 Jun 01     17:55-18:15    Quaoar      1000   41.3   2UCAC 26259184    17 32 11.415  -15 24 30.06  13.5
2012 Jun 04     01:10-01:30    1996 GQ21   302    14.0   2UCAC 28386513    15 19 25.744  -09 50 28.94  14.8
2012 Oct 07     04:30-04:45    2002 AW197  724    30.2   2UCAC 32701289    09 34 01.570  +02 31 43.27  14.7
2012 Oct 28     19:50-21:25    Varuna      603   144.8   2UCAC 41016878    07 53 29.509  +26 05 51.86  13.4
2013 Jan 29     06:10-06:30    Varuna      603    24.2   2UCAC 41187592    07 47 50.499  +26 31 18.27  14.8
2013 Feb 18     11:40-12:50    Chaos       347    93.9   2UCAC 41496855    04 49 59.491  +27 50 52.79  14.4
2013 May 28     18:35-18:55    1999 KR16   229    10.1   2UCAC 29288802    15 13 08.695  -07 07 13.91  11.7
2013 Jun 02     22:20-22:35    Ixion       759    30.4   2UCAC 21139770    17 14 19.280  -26 08 23.27  14.4
2013 Oct 23     13:20-13:35    Ixion       759    31.3   2UCAC 21137306    17 10 57.419  -26 07 23.91  14.2
2013 Dec 06     09:10-09:20    Huya        380    11.0   TYC 5025-00448-1  15 39 28.493  -05 37 11.16  10.1
2013 Dec 20     14:05-14:40    2002 TX300  724    55.4   TYC 2276-01081-1  00 52 46.890  +31 03 39.18  11.4
2013 Dec 27     17:20-17:50    1995 SM55   363    23.4   2UCAC 42006839    02 40 01.674  +29 27 23.90  13.3
2014 Jan 24     01:35-01:50    Quaoar      1000   34.9   2UCAC 26041319    17 43 03.235  -15 45 10.63  14.6
2014 Mar 01     16:30-16:45    2004 DW     1200   45.9   TYC 5476-00882-1  09 58 22.549  -08 16 55.27  12.1
2014 Mar 10     15:25-15:55    1997 CQ29   158     6.3   2UCAC 32542503    11 58 36.918  +02 28 27.68  11.9
2014 Apr 15     01:50-02:00    1996 TL66   275     9.9   2UCAC 34961430    04 26 07.463  +09 23 13.78  14.7
2014 Jul 24     14:20-15:10    Huya        380    51.0   TYC 5024-00589-1  15 36 41.149  -04 36 29.35  11.7
2014 Aug 14     21:40-22:40    1996 GQ21   302    48.1   HIP   75882       15 29 58.355  -10 03 19.73   9.1
2014 Sep 13     08:35-09:20    2002 VE95   288    27.1   2UCAC 34784210    05 31 16.439  +08 55 17.49  12.3
2014 Oct 20     14:45-15:25    1999 RZ253  219    12.4   2UCAC 29941401    23 09 49.521  -05 59 47.15  14.5
2014 Nov 08     22:00-22:15    Quaoar      1000   35.4   2UCAC 26039148    17 40 27.392  -15 41 59.91  13.9
```